\documentclass[aps,pra,showpacs,preprintnumbers,twocolumn,superscriptaddress,floatfix,nofootinbib]{revtex4-2}

\usepackage[linesnumbered,lined,boxed,commentsnumbered,ruled,vlined]{algorithm2e}
\usepackage{algpseudocode}
\usepackage{amsfonts}
\usepackage{amsmath}
\usepackage{amssymb}
\usepackage{amsthm}
\usepackage{bm,bbm}
\usepackage{caption}
\usepackage{comment}
\usepackage{dcolumn}
\usepackage{epstopdf}
\usepackage{float}
\usepackage{graphicx}
\usepackage[hypertexnames=false]{hyperref}
\usepackage{cleveref}
\usepackage{mathrsfs}
\usepackage{mathtools}
\usepackage{multirow}
\usepackage{physics}
\usepackage{ragged2e}
\usepackage{scalerel}
\usepackage{subfigure}
\usepackage{tabularx}
\usepackage{tikz}
\usepackage{url}
\usepackage{utfsym}
\usepackage{verbatim}
\usepackage{xcolor}
\usepackage{xkcdcolors}
\usepackage[normalem]{ulem}
\usepackage{lineno}
\DeclareCaptionJustification{justified}{\justifying}
\hypersetup{colorlinks=true, citecolor=orange, urlcolor=blue, linkcolor=magenta}

\definecolor{cel}{rgb}{0.0,0.53,0.74}
\definecolor{green}{rgb}{0.0,0.5,0.0}

\definecolor{colSDRG}{RGB}{51,117,56}
\definecolor{colSDCM}{RGB}{194,106,119}
\definecolor{colRec}{RGB}{230,158,0}
\definecolor{colpp}{RGB}{87,181,232}
\definecolor{colnn}{RGB}{0,158,115}
\definecolor{colpn}{RGB}{204,120,166}

\begin{document}

\title{Multiscale Renormalization of Weighted Networks}


\author{Mattia Marzi}
\email{mattia.marzi@imtlucca.it}
\affiliation{IMT School for Advanced Studies, P.zza San Francesco 19, 55100 Lucca (Italy)}
\affiliation{Lorentz Institute for Theoretical Physics, University of Leiden, Einsteinweg 55, 2333 CC Leiden (The Netherlands)}
\affiliation{Statistics Netherlands, Henri Faasdreef 312, 2492 JP Den Haag (the Netherlands)}
\affiliation{INdAM-GNAMPA Istituto Nazionale di Alta Matematica `Francesco Severi', P.le Aldo Moro 5, 00185 Rome (Italy)}
\author{Frank P. Pijpers}
\affiliation{Statistics Netherlands, Henri Faasdreef 312, 2492 JP Den Haag (the Netherlands)}
\affiliation{Korteweg - de Vries Institute for Mathematics, University of Amsterdam, Amsterdam (the Netherlands)}
\author{Diego Garlaschelli}
\affiliation{IMT School for Advanced Studies, P.zza San Francesco 19, 55100 Lucca (Italy)}
\affiliation{Lorentz Institute for Theoretical Physics, University of Leiden, Einsteinweg 55, 2333 CC Leiden (The Netherlands)}
\affiliation{INdAM-GNAMPA Istituto Nazionale di Alta Matematica `Francesco Severi', P.le Aldo Moro 5, 00185 Rome (Italy)}

\date{\today}

\begin{abstract}
Recent progress in network renormalization has identified various ways to transform network representations consistently across resolution levels. 
In particular, the renormalization flow describing how random graph models transform under node aggregation has identified a fixed point corresponding to an aggregation-invariant model that preserves the form of the connection probability across arbitrary scales. 
While this probabilistic multiscale approach has successfully been used to model real-world networks, construct latent-space node embeddings, and reconstruct finer-grained versions of aggregate network data in a principled resolution-invariant way, it has so far been restricted to binary graphs. 
Here we reformulate multiscale network renormalization for weighted random graphs.
Via a generating-function approach, we identify a weighted multiscale model whose full probability law, encoding both link creation and weight assignment, has an invariant compound-Poisson form under arbitrary node aggregation. 
The renormalization rules that transform parameters across aggregation levels allow the ensemble to be calibrated at one level and applied at another, enabling consistent modeling and reconstruction of weighted networks across arbitrary scales.
\end{abstract}

\maketitle

\makeatletter
\addtocontents{toc}{\protect\tocdepth@munge}
\makeatother
The representation of the internal architecture of all physical systems depends on the resolution at which properties are observed. 
In spatially embedded systems, geometric coordinates provide a natural way to change the resolution level, 
allowing consistent mappings across scales that lie at the foundation of the renormalization group (RG)~\cite{kadanoff1966scaling,wilson1971renormalizationI,wilson1971renormalizationII}. 
By contrast, for complex networks with no explicit spatial embedding, multiple renormalization schemes exist~\cite{Gabrielli2025,Song2005SelfSimilarity,Radicchi2009RenormalizationFlows,GarciaPerez2018MultiscaleUnfolding,Villegas2023LaplacianRG}, resulting in non-unique representations of the same system across different scales. 

The Multiscale Network Renormalization approach~\cite{Garuccio2023} has been introduced as a way to design random graph ensembles that can provide an invariant model of a network,  under arbitrary aggregations of nodes. It seeks a probability distribution over graphs that is invariant under node aggregation, thereby representing a nontrivial fixed point of an RG flow in the space of graph ensembles.
The resulting \textit{MultiScale Model} (MSM) comes with global parameters that are resolution-invariant and node variables (also called `fitness' variables) that are additive upon node aggregation.
It successfully replicates, at multiple hierarchical levels, the properties of several real-world networks~\cite{Garuccio2023,IalongoBangmaJansenGarlaschelli2024,MiloccoJansenGarlaschelli2024} and lends itself naturally to the renormalization of directed graphs~\cite{LalliGarlaschelli2024}. 
It also allows one to infer the structural properties of a network at a hierarchical level that is different from the one at which empirical observations are available~\cite{IalongoBangmaJansenGarlaschelli2024}.

Moreover, the approach can be applied to Machine Learning algorithms that take a graph as input and encode its structure onto latent output vectors that represent nodes in an abstract space~\cite{MiloccoJansenGarlaschelli2024,MiloccoJansenGarlaschelli2025}. Under arbitrary coarse-grainings of the input graph, the multiscale method ensures statistical consistency of the embedding vector of a block-node with the sum of the embedding vectors of its constituent nodes. 
This guarantee enables the interpretable application of the basic properties of vector spaces (i.e. sum of vectors and multiplication of a vector by a scalar) to the latent space where node embeddings are identified.  
Several key network properties, including a large number of triangles, are successfully replicated already from embeddings of very low dimensionality, allowing for the generation of faithful replicas of the original networks at arbitrary resolution~\cite{MiloccoJansenGarlaschelli2024,MiloccoJansenGarlaschelli2025}.

Finally, an annealed version of the MSM (where the fitness is interpreted as a random variable) leads to infinite-mean node variables and  spontaneously replicates several real-world network properties, such as power-law degree distributions~\cite{Garuccio2023,avena2026inhomogeneous}, disassortativity profiles~\cite{Garuccio2023}, and the surprising coexistence of sparsity and clustering~\cite{Garuccio2023,catanzaro2026clustering}, without the need to introduce geometric constructions or edge dependencies. 
Since node aggregation invariance is a form of discrete scale invariance, several unique properties emerge in the spectrum of the adjacency matrices generated by the model, such as log-periodicity and complex scaling exponents~\cite{catanzaro2025spectra}.
Recent research has shown that this annealed version of the MSM represents a heterogeneous attractor for an RG flow in the space of `heavy-tailed' random graph models, in contrast with the Erd\H{o}s-R\'enyi model which stands as a homogeneous attractor for `light-tailed' random graphs~\cite{avena2026renormalisation}. This finding provides a natural mechanism through which the MSM spontaneously emerges as an effective model for a large class of microscopic network formation mechanisms.

\begin{figure*}[t!]
\centering
\includegraphics[width=0.95\textwidth]{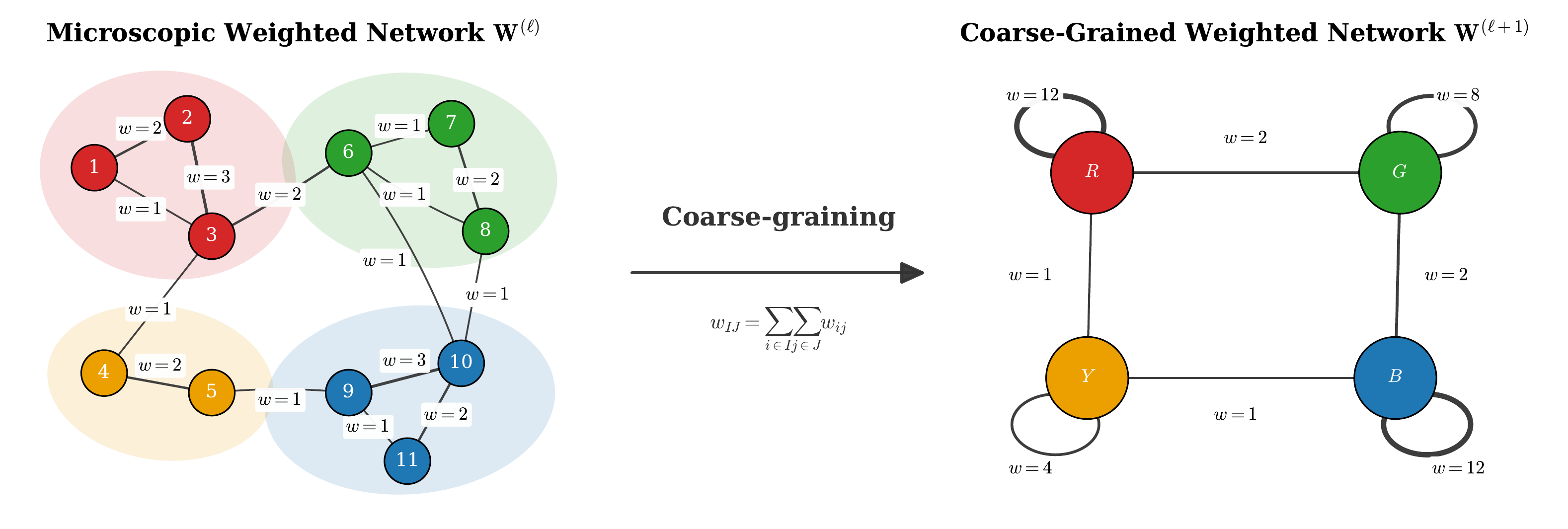}
\caption{{\textbf{Weighted coarse-graining under node aggregation.} Microscopic nodes at layer $\ell$ are grouped into blocks, which define the macro-nodes at layer $\ell+1$. Two macro-nodes are connected if at least one pair of microscopic nodes across the corresponding blocks is connected at layer $\ell$. The weight between coarse-grained nodes is the sum of all microscopic weights across them, while microscopic weights internal to the same block become weighted self-loops.}}
\label{fig:renormalization_rules}
\end{figure*}

{Weighted network renormalization has been addressed through geometric~\cite{Chen2023WeightedGR,Zheng2024WeightedGR} and Laplacian approaches~\cite{Caldarelli2024LaplacianRG}. Here, we extend the MSM to weighted networks.}
Our main result is a weighted MSM (wMSM) generalizing  the aggregation-invariant construction from the level of purely binary topology to that of (non-negative integer-valued) weighted adjacency matrices. We identify an invariant compound-Poisson specification {for dyads between distinct blocks} that, crucially, separates the control of link probabilities (which should not depend on the arbitrary units chosen for the link weights) from that of the link weights (whose units should be assigned independently of the bare topology). 
This yields a scheme in which a global density parameter and a global link weight parameter can be calibrated separately.
In a companion paper~\cite{Marzi2026WeightedNetworksAcrossScales}, we show a detailed application of the wMSM to real-world networks, where we confirm its cross-scale reconstruction capabilities.
Specifically, the global model parameters can be calibrated at an observed aggregate layer and then transferred unchanged to finer layers, to provide an inferential framework that is consistent upon both coarse- and fine-graining {for inter-block distributions and full-matrix weight expectations; the diagonal distribution requires a separate treatment. More details are found in Supplementary Information (SI)}. 

Before describing the wMSM, we briefly summarize the salient features of the binary MSM~\cite{Garuccio2023} (see also SI). Consider a partition of nodes into disjoint blocks. If microscopic links are independent and drawn with connection probabilities $p_{ij}$, the probability that at least one connection exists between block $I$ and block $J$, with $I\neq J$, obeys
\begin{equation}
p_{IJ}=1-\prod_{i\in I}\prod_{j\in J}(1-p_{ij}),\qquad{I\neq J}.
\end{equation}
A random graph model is said to be aggregation-invariant if the coarse-grained probability $p_{IJ}$ belongs to the same parametric family as $p_{ij}$, with block-level parameters obtained through deterministic aggregation rules~\cite{Garuccio2023,LalliGarlaschelli2024,IalongoBangmaJansenGarlaschelli2024,Gabrielli2025}.
Requiring closure of the family under the above coarse-graining operation uniquely selects, for independent dyads between distinct blocks, the following exponential functional form of the MSM~\cite{Garuccio2023}:
\begin{equation}
p_{ij}=1-\exp(-\delta x_i x_j),\qquad{i\neq j},\label{eq_p}
\end{equation}
where $\delta>0$ is a global invariant parameter and $x_i\ge0$ is a local parameter tuning the likelihood of node $i$ to form connections, transforming additively under aggregation:
\begin{equation}
x_I=\sum_{i\in I}x_i.
\end{equation}
At the block level, one then obtains
\begin{equation}
p_{IJ}=1-\exp(-\delta x_I x_J),\qquad{I\neq J},
\end{equation}
so that aggregating the network after specifying the model yields the same functional form as specifying the model directly at the aggregated level, given the block parameters $\{x_I\}$. 
The full analytical derivation of this result is reported in SI.
The extension to directed networks is possible and is obtained by assigning separate additive fields to outgoing and incoming node activity, and possibly to reciprocated in-out activity as well~\cite{LalliGarlaschelli2024}.

The aggregation-invariance property becomes particularly relevant in empirical settings where certain constraints, for instance the overall link density, are observable only at aggregated resolutions. In such cases, one may wish to calibrate the model at a coarse-grained level and then infer microscopic structure consistently with those constraints. If the functional form is not preserved under aggregation, this procedure becomes ill-defined: parameters must be re-estimated at each resolution, and the resulting microscopic network is no longer guaranteed to reproduce the observed aggregated constraints.

We now extend the scale-invariant construction to \textit{integer-valued} weights $w_{ij}\in\mathbb N$.
As a preliminary observation, we note that in the {weighted} case we still want the purely binary projection of the graph to obey the same scale-invariance requirement as in the ordinary MSM: we therefore require that the expected value of the binary projection of the weighted adjacency matrix entry $w_{ij}$ is still governed by Eq.~\eqref{eq_p}. 
We use uppercase letters for adjacency and weight matrices and lowercase letters for their entries; model averages are indicated by explicit expectation operators. Additionally, we note that a convenient (although not necessary) choice for the node parameter $x_i$ is (up to a proportionality constant) the {observed} strength $s_i$ of the same node, i.e.
\begin{equation}
x_i=s_i,
\end{equation}
yielding the following specification of the binary MSM:
\begin{equation}
p^{\mathrm{MSM}}_{ij}=1-\exp\!\big(-\delta\,s_i s_j\big),\qquad{i\neq j}.
\end{equation}
Since strengths are additive under aggregation, block parameters automatically satisfy $x_I=s_I=\sum_{i\in I}s_i$, ensuring multiscale consistency.

For the weighted model, we also require closure under coarse graining by summation:
\begin{equation}
w_{IJ}=\sum_{i\in I}\sum_{j\in J}w_{ij},
\qquad I\neq J.
\end{equation}
For diagonal blocks, $w_{II}$ represents the total internal weight of block $I$ and is obtained by summing the microscopic weights whose endpoints both belong to $I$. {Thus $w_{II}=\sum_{i\in I}w_{ii}+2\sum_{i\in I}\sum_{j\in I,\,j>i}w_{ij}$, which preserves row-strength additivity. The diagonal PGF and its admissibility conditions are derived in SI.} {The corresponding renormalization rule is illustrated in Figure~\ref{fig:renormalization_rules}: microscopic weights are summed across pairs of blocks, while internal microscopic weights become diagonal block weights at the coarse-grained scale.} Under conditional dyadic independence, the natural object controlling sums of integer random variables is the Probability Generating Function (PGF) $G_{ij}(z)=E[z^{w_{ij}}]$, $z\in[0,1]$. This function encodes the weight distribution, since $G_{ij}(z)=\sum_{n=0}^{\infty}P(w_{ij}=n)z^n$. For disjoint blocks $I,J$ one then has
\begin{equation}
G_{IJ}(z)=E[z^{w_{IJ}}]=\prod_{i\in I}\prod_{j\in J}G_{ij}(z),\qquad{I\neq J}.
\end{equation}
Requiring that the family remains closed under arbitrary aggregation with additive node fields $x_I=\sum_{i\in I}x_i$ selects the exponential bilinear structure
\begin{equation}
G_{ij}(z)=\exp\!\big(-x_i x_j\,\psi(z)\big),\qquad{i\neq j}
\end{equation}
for some nonnegative function $\psi(z)$. Normalization of the probability distribution requires $G_{ij}(1)=E[1^{w_{ij}}]=1$. Substituting into $G_{ij}(z)=\exp(-x_i x_j\,\psi(z))$ yields $1=\exp(-x_i x_j\,\psi(1))$, which implies $\psi(1)=0$. {Requiring this PGF to remain valid for arbitrary nonnegative node fields implies infinite divisibility on the nonnegative integers, hence a compound-Poisson representation by Feller's characterization~\cite{Feller1968,OspinaGerber1987} (see also SI).}

To turn the scale-invariant PGF representation into an explicit probabilistic model, one must still specify a concrete dyadic weight distribution. We adopt the compound-Poisson construction~\cite{Feller1971,JohnsonKempKotz2005}, representing each dyadic volume as the sum of a random number of positive integer contributions,
\begin{equation}
w_{ij}=\sum_{r=1}^{K_{ij}}X_{ij,r},\qquad K_{ij}\sim\mathrm{Poisson}(\lambda_{ij}),
\end{equation}
where $K_{ij}$ counts the positive contributions to the dyadic weight, and the marks $X_{ij,r}\in\{1,2,\dots\}$ specify their sizes. The marks are i.i.d. and independent of $K_{ij}$. Denoting their common PGF by $h(z)=E[z^X]$, independence gives $E[z^{w_{ij}}\mid K_{ij}=k]=[h(z)]^k$. Averaging over the Poisson count then yields
\begin{equation}
G_{ij}(z)=E[h(z)^{K_{ij}}]=\exp\{\lambda_{ij}[h(z)-1]\},
\end{equation}
which belongs to the scale invariant family {between distinct blocks}, since Poisson intensities add under aggregation.

\begin{figure*}[t!]
\centering
\includegraphics[width=0.95\textwidth]{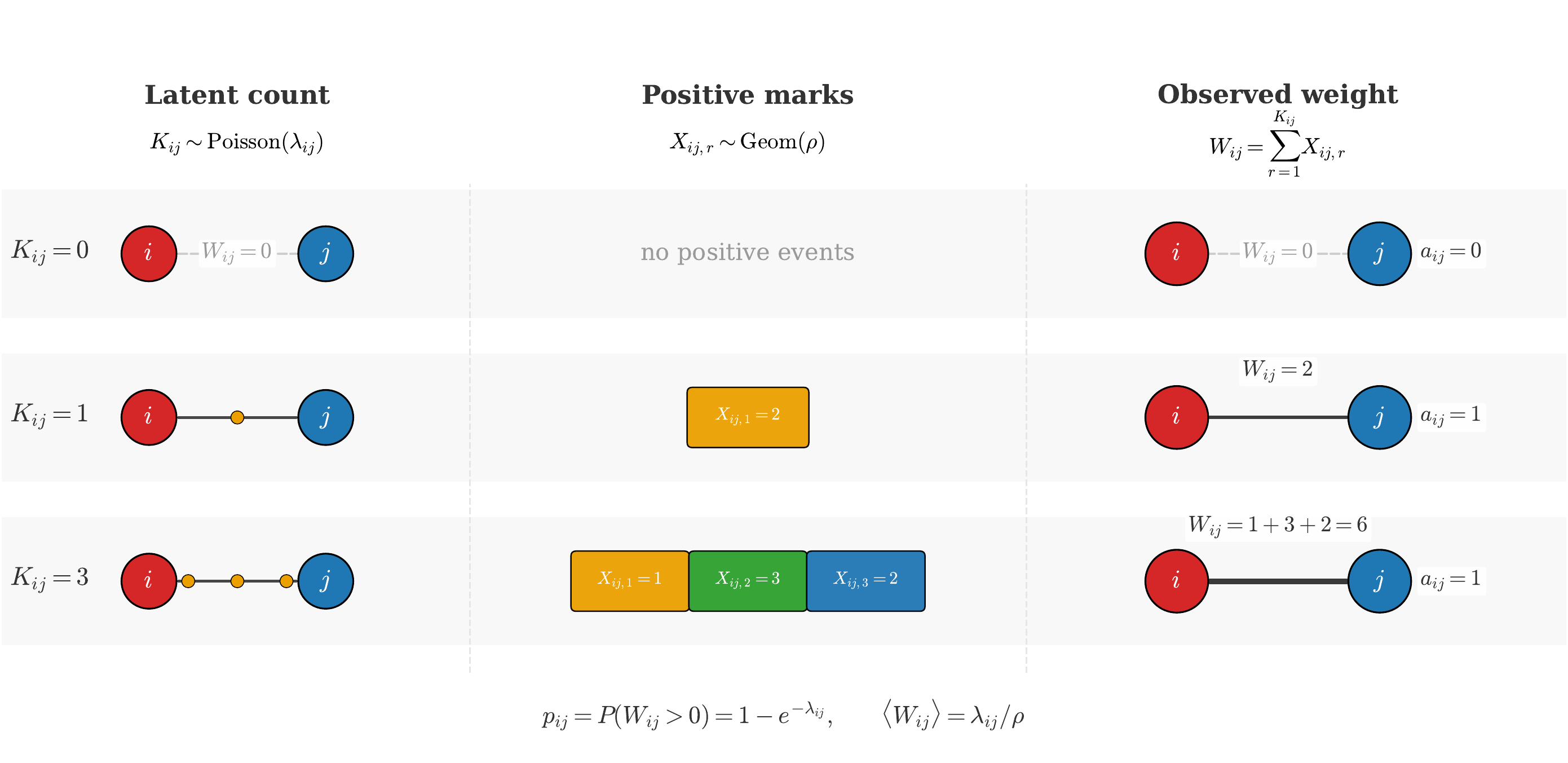}
\caption{{\textbf{Compound-Poisson mechanism of the weighted MultiScale Model}. For each dyad $(i,j)$, a Poisson latent count $K_{ij}$ determines the number of positive exchange events. If $K_{ij}=0$, no link is observed and $w_{ij}=0$; if $K_{ij}\ge1$, the dyad is connected. Each event contributes a positive geometric mark $X_{ij,r}$ with mean $E[X]=1/\rho$, and the dyadic weight is the sum of all marks, $w_{ij}=\sum_{r=1}^{K_{ij}}X_{ij,r}$. Thus $\lambda_{ij}$ controls link formation, through $P(w_{ij}>0)=1-\exp(-\lambda_{ij})$, whereas $\rho$ controls the typical size of positive weight contributions.}}
\label{fig:msm_mechanism}
\end{figure*}

To obtain a weighted analogue consistent with the binary MSM, we drop dyadic frictions and set
\begin{equation}
\lambda_{ij}=\delta\,s_i s_j,\qquad{i\neq j}.
\end{equation}
With this choice, the compound-Poisson PGF has the exponential bilinear form introduced above, with $x_i=s_i$ and $\psi(z)=\delta[1-h(z)]$.
Together with geometric marks {of parameter $0<\rho\le1$},
\begin{equation}
P(X=n)=\rho(1-\rho)^{n-1},\qquad n\ge1,
\end{equation}
this defines the wMSM as the compound-Poisson law for
\begin{equation}
w_{ij}=\sum_{r=1}^{K_{ij}}X_{ij,r},\qquad K_{ij}\sim\mathrm{Poisson}(\lambda_{ij}).
\end{equation}
The resulting dyadic distribution reads, for $n=0$,
\begin{equation}
P^{\mathrm{wMSM}}(w_{ij}=0)=\exp\!\Big(-\delta\,s_i s_j\Big),\qquad{i\neq j},
\end{equation}
and, for $n\ge1$,
\begin{align}
&P^{\mathrm{wMSM}}(w_{ij}=n)=\exp\Big(-\delta\,s_i s_j\Big)\nonumber\\
&\cdot\sum_{k=1}^{n}\frac{\left(\delta\,s_i s_j\right)^k}{k!}\binom{n-1}{k-1}\rho^k(1-\rho)^{n-k},\qquad{i\neq j}.
\end{align}
From the atom at zero we recover the link probabilities,
\begin{align}
p_{ij}^{\mathrm{wMSM}}
&=P^{\mathrm{wMSM}}(w_{ij}>0)=1-P^{\mathrm{wMSM}}(w_{ij}=0)\nonumber\\
&=1-\exp\!\Big(-\delta\,s_i s_j\Big),\qquad{i\neq j},
\end{align}
while expected weights are found conditioning on $K_{ij}$:
\begin{align}
E^{\mathrm{wMSM}}[w_{ij}]
&=E\!\left[E\!\left[\sum_{r=1}^{K_{ij}}X_{ij,r}\,\Big|\,K_{ij}\right]\right]=E\!\left[K_{ij}\,E[X]\right]\nonumber\\
&=E[K_{ij}]\,E[X]=\frac{\delta}{\rho}\,s_i s_j,\qquad{i\neq j}.
\end{align}

{For the reconstruction prescription used in the companion paper, the same dyadic law is also assigned to retained diagonal entries. This choice preserves full-matrix weight expectations, while exact distributional closure is restricted to distinct blocks (see also SI).} Let $W=\sum_i s_i$ denote the observed total strength in the matrix convention adopted throughout the paper. Using $E[X]=1/\rho$, the expected strength factorizes as
\begin{equation}
E^\mathrm{wMSM}[s_i]
=\sum_{j}E^{\mathrm{wMSM}}[w_{ij}] =\frac{\delta}{\rho}\,s_i\sum_{j}s_j=\frac{\delta}{\rho}\,s_i W\nonumber
\end{equation}
%
%
So, choosing
\begin{equation}
\rho=\delta\,W
\end{equation}
enforces $E^\mathrm{wMSM}[s_i]=s_i$ for all nodes whenever the full dyadic support at the chosen scale is retained{, provided $0<\delta W\le1$}. The parameter $\delta$ controls the binary projection through $p_{ij}^{\mathrm{wMSM}}=1-\exp(-\delta s_i s_j)$, while $\rho$ rescales positive weights without affecting link probabilities. This additional degree of freedom is necessary in our reconstruction setting, where topological sparsity and aggregate weight constraints must be matched independently. 

The compound-Poisson mechanism is illustrated in Figure~\ref{fig:msm_mechanism}. Conceptually, the construction separates two aspects of a weighted interaction that would otherwise be tied together. The latent count $K_{ij}$ represents the number of positive exchange events contributing to the dyadic weight within the observation window. If $K_{ij}=0$, the dyad has zero weight and no link is observed; if $K_{ij}\ge1$, a link is present. Therefore the intensity $\lambda_{ij}=\delta s_i s_j$ controls the binary projection, since $p_{ij}^{\mathrm{wMSM}}=P(K_{ij}\ge1)=1-\exp(-\lambda_{ij})$. Conditional on these latent events, the marks $X_{ij,r}$ determine how much weight each event contributes. Each contribution starts with one weight unit and, after each unit, stops with probability $\rho$ or adds another unit with probability $1-\rho$. A size $n$ therefore corresponds to $n-1$ continuations followed by one stop, yielding the geometric law. The parameter $\rho$ controls mean size, since $E[X]=1/\rho$, and therefore rescales expected weights without changing link probabilities. In economic applications, this should be understood as a discretized representation of exchanged value, with $\rho$ absorbing the conversion between the chosen weight unit and the average size of a latent exchange contribution. The negative-binomial term in the dyadic distribution arises because, conditional on $K_{ij}=k$, the observed weight $w_{ij}=n$ is obtained as the sum of $k$ positive geometric contributions.

{Self-loops require a separate treatment: exact closure of their distribution under aggregation is not guaranteed by the off-diagonal construction. For strength-based reconstruction, we adopt a diagonal prescription that preserves all expected node strengths with the single normalization $\rho=\delta W$ when diagonal entries are retained, while relaxing exact closure of the diagonal distribution. Excluding diagonal entries instead introduces a correction to the expected strengths, as in other strength-driven reconstruction methods~\cite{ReconstructingFirmLevelInteractionsDutchInputOutputNetwork2022,CReM2020}. The diagonal constructions, their admissibility conditions, and the reconstruction prescription are detailed in SI.}

Closure under coarse graining follows from additivity of Poisson intensities. In particular, for any two distinct blocks $I,J$ one has $s_I=\sum_{i\in I}s_i$ and $s_J=\sum_{j\in J}s_j$, hence $\lambda_{IJ}=\sum_{i\in I}\sum_{j\in J}\lambda_{ij}=\delta\,s_I s_J$ and the wMSM remains exactly multiscale consistent.
The model therefore combines two ingredients. The binary projection follows the scale-invariant exponential form, which is closed under aggregation when node variables are additive. The weighted layer is obtained through a compound-Poisson construction, which separates the control of link probabilities from the normalization of expected weights. 
This separation is important: it can match the chosen scale of weights   while leaving the topology independent of such scale.

In the companion paper~\cite{Marzi2026WeightedNetworksAcrossScales}, we apply the wMSM to two real-world systems that feature natural -- but distinct -- hierarchical aggregation schemes: the International Trade Network, that can be coarse-grained following geographic proximity, and a nation-wide production network, that can be coarse-grained following sectoral similarity.
We show that the transferred coarse-scale calibration remains informative at the microscopic level. In particular, the method preserves strength information, while accurately predicting the number of fine-scale links and local structural profiles that are not imposed as constraints. When compared with a fine-scale weighted reconstruction calibrated directly at the target resolution and sharing the same unconditional gravity expectation once strengths are fixed, the scale-invariant approach systematically improves the selectivity of the inferred binary support, increasing precision, specificity and accuracy while keeping sensitivity nearly unchanged. 

Network modeling is typically formulated at a fixed resolution~\cite{marzi2026reproducing}, including when information is transferred across successive snapshots~\cite{marzi2026Bayesian}: constraints are observed at a given scale and the ensemble is calibrated at that same scale. This assumption is restrictive in many empirical settings, where the available information is aggregated while the relevant interactions occur at a finer level. Moreover, a change of resolution in the definition of the original system implies in general an undesired departure of the  graph ensemble used to describe the system from its original functional form. Here we have introduced the wMSM as a reconstruction framework designed precisely for this situation. Its defining property is that the same probabilistic form is preserved under coarse graining, so that a parameter calibrated on an observed aggregate layer can be transferred to finer layers without refitting. 

\section*{ACKNOWLEDGMENTS}

This publication is part of the projects ``Network renormalization: from theoretical physics to the resilience of societies’’ with file number NWA.1418.24.029 of the research programme NWA L3 - Innovative projects within routes 2024, which is (partly) financed by the Dutch Research Council (NWO) under the grant \url{https://doi.org/10.61686/AOIJP05368}, and ``Redefining renormalization for complex networks’’ with file number OCENW.M.24.039 of the research programme Open Competition Domain Science Package 24-1, which is (partly) financed by the Dutch Research Council (NWO) under the grant \url{https://doi.org/10.61686/PBSEC42210}.

\bibliography{references}

\begin{thebibliography}{31}%
\makeatletter
\providecommand \@ifxundefined [1]{%
 \@ifx{#1\undefined}
}%
\providecommand \@ifnum [1]{%
 \ifnum #1\expandafter \@firstoftwo
 \else \expandafter \@secondoftwo
 \fi
}%
\providecommand \@ifx [1]{%
 \ifx #1\expandafter \@firstoftwo
 \else \expandafter \@secondoftwo
 \fi
}%
\providecommand \natexlab [1]{#1}%
\providecommand \enquote  [1]{``#1''}%
\providecommand \bibnamefont  [1]{#1}%
\providecommand \bibfnamefont [1]{#1}%
\providecommand \citenamefont [1]{#1}%
\providecommand \href@noop [0]{\@secondoftwo}%
\providecommand \href [0]{\begingroup \@sanitize@url \@href}%
\providecommand \@href[1]{\@@startlink{#1}\@@href}%
\providecommand \@@href[1]{\endgroup#1\@@endlink}%
\providecommand \@sanitize@url [0]{\catcode `\\12\catcode `\$12\catcode `\&12\catcode `\#12\catcode `\^12\catcode `\_12\catcode `\%12\relax}%
\providecommand \@@startlink[1]{}%
\providecommand \@@endlink[0]{}%
\providecommand \url  [0]{\begingroup\@sanitize@url \@url }%
\providecommand \@url [1]{\endgroup\@href {#1}{\urlprefix }}%
\providecommand \urlprefix  [0]{URL }%
\providecommand \Eprint [0]{\href }%
\providecommand \doibase [0]{https://doi.org/}%
\providecommand \selectlanguage [0]{\@gobble}%
\providecommand \bibinfo  [0]{\@secondoftwo}%
\providecommand \bibfield  [0]{\@secondoftwo}%
\providecommand \translation [1]{[#1]}%
\providecommand \BibitemOpen [0]{}%
\providecommand \bibitemStop [0]{}%
\providecommand \bibitemNoStop [0]{.\EOS\space}%
\providecommand \EOS [0]{\spacefactor3000\relax}%
\providecommand \BibitemShut  [1]{\csname bibitem#1\endcsname}%
\let\auto@bib@innerbib\@empty
\bibitem [{\citenamefont {Kadanoff}(1966)}]{kadanoff1966scaling}%
  \BibitemOpen
  \bibfield  {author} {\bibinfo {author} {\bibfnamefont {L.~P.}\ \bibnamefont {Kadanoff}},\ }\bibfield  {title} {\bibinfo {title} {Scaling laws for ising models near $t_c$},\ }\href {https://doi.org/10.1103/PhysicsPhysiqueFizika.2.263} {\bibfield  {journal} {\bibinfo  {journal} {Physics Physique Fizika}\ }\textbf {\bibinfo {volume} {2}},\ \bibinfo {pages} {263} (\bibinfo {year} {1966})}\BibitemShut {NoStop}%
\bibitem [{\citenamefont {Wilson}(1971{\natexlab{a}})}]{wilson1971renormalizationI}%
  \BibitemOpen
  \bibfield  {author} {\bibinfo {author} {\bibfnamefont {K.~G.}\ \bibnamefont {Wilson}},\ }\bibfield  {title} {\bibinfo {title} {Renormalization group and critical phenomena. i. renormalization group and the kadanoff scaling picture},\ }\href {https://doi.org/10.1103/PhysRevB.4.3174} {\bibfield  {journal} {\bibinfo  {journal} {Physical Review B}\ }\textbf {\bibinfo {volume} {4}},\ \bibinfo {pages} {3174} (\bibinfo {year} {1971}{\natexlab{a}})}\BibitemShut {NoStop}%
\bibitem [{\citenamefont {Wilson}(1971{\natexlab{b}})}]{wilson1971renormalizationII}%
  \BibitemOpen
  \bibfield  {author} {\bibinfo {author} {\bibfnamefont {K.~G.}\ \bibnamefont {Wilson}},\ }\bibfield  {title} {\bibinfo {title} {Renormalization group and critical phenomena. ii. phase-space cell analysis of critical behavior},\ }\href {https://doi.org/10.1103/PhysRevB.4.3184} {\bibfield  {journal} {\bibinfo  {journal} {Physical Review B}\ }\textbf {\bibinfo {volume} {4}},\ \bibinfo {pages} {3184} (\bibinfo {year} {1971}{\natexlab{b}})}\BibitemShut {NoStop}%
\bibitem [{\citenamefont {Gabrielli}\ \emph {et~al.}(2025)\citenamefont {Gabrielli}, \citenamefont {Garlaschelli}, \citenamefont {Patil},\ and\ \citenamefont {Serrano}}]{Gabrielli2025}%
  \BibitemOpen
  \bibfield  {author} {\bibinfo {author} {\bibfnamefont {A.}~\bibnamefont {Gabrielli}}, \bibinfo {author} {\bibfnamefont {D.}~\bibnamefont {Garlaschelli}}, \bibinfo {author} {\bibfnamefont {S.~P.}\ \bibnamefont {Patil}},\ and\ \bibinfo {author} {\bibfnamefont {M.~{\'A}.}\ \bibnamefont {Serrano}},\ }\bibfield  {title} {\bibinfo {title} {Network renormalization},\ }\href {https://doi.org/10.1038/s42254-025-00817-5} {\bibfield  {journal} {\bibinfo  {journal} {Nat. Rev. Phys.}\ }\textbf {\bibinfo {volume} {7}},\ \bibinfo {pages} {203} (\bibinfo {year} {2025})}\BibitemShut {NoStop}%
\bibitem [{\citenamefont {Song}\ \emph {et~al.}(2005)\citenamefont {Song}, \citenamefont {Havlin},\ and\ \citenamefont {Makse}}]{Song2005SelfSimilarity}%
  \BibitemOpen
  \bibfield  {author} {\bibinfo {author} {\bibfnamefont {C.}~\bibnamefont {Song}}, \bibinfo {author} {\bibfnamefont {S.}~\bibnamefont {Havlin}},\ and\ \bibinfo {author} {\bibfnamefont {H.~A.}\ \bibnamefont {Makse}},\ }\bibfield  {title} {\bibinfo {title} {Self-similarity of complex networks},\ }\href {https://doi.org/10.1038/nature03248} {\bibfield  {journal} {\bibinfo  {journal} {Nature}\ }\textbf {\bibinfo {volume} {433}},\ \bibinfo {pages} {392} (\bibinfo {year} {2005})}\BibitemShut {NoStop}%
\bibitem [{\citenamefont {Radicchi}\ \emph {et~al.}(2009)\citenamefont {Radicchi}, \citenamefont {Ramasco}, \citenamefont {Barrat},\ and\ \citenamefont {Fortunato}}]{Radicchi2009RenormalizationFlows}%
  \BibitemOpen
  \bibfield  {author} {\bibinfo {author} {\bibfnamefont {F.}~\bibnamefont {Radicchi}}, \bibinfo {author} {\bibfnamefont {J.~J.}\ \bibnamefont {Ramasco}}, \bibinfo {author} {\bibfnamefont {A.}~\bibnamefont {Barrat}},\ and\ \bibinfo {author} {\bibfnamefont {S.}~\bibnamefont {Fortunato}},\ }\bibfield  {title} {\bibinfo {title} {Renormalization flows in complex networks},\ }\href {https://doi.org/10.1103/PhysRevE.79.026104} {\bibfield  {journal} {\bibinfo  {journal} {Physical Review E}\ }\textbf {\bibinfo {volume} {79}},\ \bibinfo {pages} {026104} (\bibinfo {year} {2009})}\BibitemShut {NoStop}%
\bibitem [{\citenamefont {Garc{\'i}a-P{\'e}rez}\ \emph {et~al.}(2018)\citenamefont {Garc{\'i}a-P{\'e}rez}, \citenamefont {Bogu{\~n}{\'a}},\ and\ \citenamefont {Serrano}}]{GarciaPerez2018MultiscaleUnfolding}%
  \BibitemOpen
  \bibfield  {author} {\bibinfo {author} {\bibfnamefont {G.}~\bibnamefont {Garc{\'i}a-P{\'e}rez}}, \bibinfo {author} {\bibfnamefont {M.}~\bibnamefont {Bogu{\~n}{\'a}}},\ and\ \bibinfo {author} {\bibfnamefont {M.~{\'A}.}\ \bibnamefont {Serrano}},\ }\bibfield  {title} {\bibinfo {title} {Multiscale unfolding of real networks by geometric renormalization},\ }\href {https://doi.org/10.1038/s41567-018-0072-5} {\bibfield  {journal} {\bibinfo  {journal} {Nature Physics}\ }\textbf {\bibinfo {volume} {14}},\ \bibinfo {pages} {583} (\bibinfo {year} {2018})}\BibitemShut {NoStop}%
\bibitem [{\citenamefont {Villegas}\ \emph {et~al.}(2023)\citenamefont {Villegas}, \citenamefont {Gili}, \citenamefont {Caldarelli},\ and\ \citenamefont {Gabrielli}}]{Villegas2023LaplacianRG}%
  \BibitemOpen
  \bibfield  {author} {\bibinfo {author} {\bibfnamefont {P.}~\bibnamefont {Villegas}}, \bibinfo {author} {\bibfnamefont {T.}~\bibnamefont {Gili}}, \bibinfo {author} {\bibfnamefont {G.}~\bibnamefont {Caldarelli}},\ and\ \bibinfo {author} {\bibfnamefont {A.}~\bibnamefont {Gabrielli}},\ }\bibfield  {title} {\bibinfo {title} {Laplacian renormalization group for heterogeneous networks},\ }\href {https://doi.org/10.1038/s41567-022-01866-8} {\bibfield  {journal} {\bibinfo  {journal} {Nature Physics}\ }\textbf {\bibinfo {volume} {19}},\ \bibinfo {pages} {445} (\bibinfo {year} {2023})}\BibitemShut {NoStop}%
\bibitem [{\citenamefont {Garuccio}\ \emph {et~al.}(2023)\citenamefont {Garuccio}, \citenamefont {Lalli},\ and\ \citenamefont {Garlaschelli}}]{Garuccio2023}%
  \BibitemOpen
  \bibfield  {author} {\bibinfo {author} {\bibfnamefont {E.}~\bibnamefont {Garuccio}}, \bibinfo {author} {\bibfnamefont {M.}~\bibnamefont {Lalli}},\ and\ \bibinfo {author} {\bibfnamefont {D.}~\bibnamefont {Garlaschelli}},\ }\bibfield  {title} {\bibinfo {title} {Multiscale network renormalization: Scale-invariance without geometry},\ }\href {https://doi.org/10.1103/PhysRevResearch.5.043101} {\bibfield  {journal} {\bibinfo  {journal} {Phys. Rev. Res.}\ }\textbf {\bibinfo {volume} {5}},\ \bibinfo {pages} {043101} (\bibinfo {year} {2023})}\BibitemShut {NoStop}%
\bibitem [{\citenamefont {Ialongo}\ \emph {et~al.}(2024)\citenamefont {Ialongo}, \citenamefont {Bangma}, \citenamefont {Jansen},\ and\ \citenamefont {Garlaschelli}}]{IalongoBangmaJansenGarlaschelli2024}%
  \BibitemOpen
  \bibfield  {author} {\bibinfo {author} {\bibfnamefont {L.~N.}\ \bibnamefont {Ialongo}}, \bibinfo {author} {\bibfnamefont {S.}~\bibnamefont {Bangma}}, \bibinfo {author} {\bibfnamefont {F.}~\bibnamefont {Jansen}},\ and\ \bibinfo {author} {\bibfnamefont {D.}~\bibnamefont {Garlaschelli}},\ }\bibfield  {title} {\bibinfo {title} {Multi-scale reconstruction of large supply networks},\ }\bibfield  {journal} {\bibinfo  {journal} {arXiv}\ }\href {https://doi.org/10.48550/arXiv.2412.16122} {10.48550/arXiv.2412.16122} (\bibinfo {year} {2024})\BibitemShut {NoStop}%
\bibitem [{\citenamefont {Milocco}\ \emph {et~al.}(2024)\citenamefont {Milocco}, \citenamefont {Jansen},\ and\ \citenamefont {Garlaschelli}}]{MiloccoJansenGarlaschelli2024}%
  \BibitemOpen
  \bibfield  {author} {\bibinfo {author} {\bibfnamefont {R.}~\bibnamefont {Milocco}}, \bibinfo {author} {\bibfnamefont {F.}~\bibnamefont {Jansen}},\ and\ \bibinfo {author} {\bibfnamefont {D.}~\bibnamefont {Garlaschelli}},\ }\bibfield  {title} {\bibinfo {title} {Multi-scale node embeddings for graph modeling and generation},\ }\bibfield  {journal} {\bibinfo  {journal} {arXiv}\ }\href {https://doi.org/10.48550/arXiv.2412.04354} {10.48550/arXiv.2412.04354} (\bibinfo {year} {2024})\BibitemShut {NoStop}%
\bibitem [{\citenamefont {Lalli}\ and\ \citenamefont {Garlaschelli}(2024)}]{LalliGarlaschelli2024}%
  \BibitemOpen
  \bibfield  {author} {\bibinfo {author} {\bibfnamefont {M.}~\bibnamefont {Lalli}}\ and\ \bibinfo {author} {\bibfnamefont {D.}~\bibnamefont {Garlaschelli}},\ }\bibfield  {title} {\bibinfo {title} {Geometry-free renormalization of directed networks: scale-invariance and reciprocity},\ }\bibfield  {journal} {\bibinfo  {journal} {arXiv}\ }\href {https://doi.org/10.48550/arXiv.2403.00235} {10.48550/arXiv.2403.00235} (\bibinfo {year} {2024})\BibitemShut {NoStop}%
\bibitem [{\citenamefont {Milocco}\ \emph {et~al.}(2025)\citenamefont {Milocco}, \citenamefont {Jansen},\ and\ \citenamefont {Garlaschelli}}]{MiloccoJansenGarlaschelli2025}%
  \BibitemOpen
  \bibfield  {author} {\bibinfo {author} {\bibfnamefont {R.}~\bibnamefont {Milocco}}, \bibinfo {author} {\bibfnamefont {F.}~\bibnamefont {Jansen}},\ and\ \bibinfo {author} {\bibfnamefont {D.}~\bibnamefont {Garlaschelli}},\ }\bibfield  {title} {\bibinfo {title} {Renormalizable graph embeddings for multi-scale network reconstruction},\ }\bibfield  {journal} {\bibinfo  {journal} {arXiv}\ }\href {https://doi.org/10.48550/arXiv.2508.20706} {10.48550/arXiv.2508.20706} (\bibinfo {year} {2025})\BibitemShut {NoStop}%
\bibitem [{\citenamefont {Avena}\ \emph {et~al.}(2026{\natexlab{a}})\citenamefont {Avena}, \citenamefont {Garlaschelli}, \citenamefont {Hazra},\ and\ \citenamefont {Lalli}}]{avena2026inhomogeneous}%
  \BibitemOpen
  \bibfield  {author} {\bibinfo {author} {\bibfnamefont {L.}~\bibnamefont {Avena}}, \bibinfo {author} {\bibfnamefont {D.}~\bibnamefont {Garlaschelli}}, \bibinfo {author} {\bibfnamefont {R.~S.}\ \bibnamefont {Hazra}},\ and\ \bibinfo {author} {\bibfnamefont {M.}~\bibnamefont {Lalli}},\ }\bibfield  {title} {\bibinfo {title} {Inhomogeneous random graphs with infinite-mean fitness variables},\ }\href {https://doi.org/10.1017/jpr.2025.10037} {\bibfield  {journal} {\bibinfo  {journal} {Journal of Applied Probability}\ }\textbf {\bibinfo {volume} {63}},\ \bibinfo {pages} {375} (\bibinfo {year} {2026}{\natexlab{a}})}\BibitemShut {NoStop}%
\bibitem [{\citenamefont {Catanzaro}\ \emph {et~al.}(2026{\natexlab{a}})\citenamefont {Catanzaro}, \citenamefont {van~der Hofstad},\ and\ \citenamefont {Garlaschelli}}]{catanzaro2026clustering}%
  \BibitemOpen
  \bibfield  {author} {\bibinfo {author} {\bibfnamefont {A.}~\bibnamefont {Catanzaro}}, \bibinfo {author} {\bibfnamefont {R.}~\bibnamefont {van~der Hofstad}},\ and\ \bibinfo {author} {\bibfnamefont {D.}~\bibnamefont {Garlaschelli}},\ }\bibfield  {title} {\bibinfo {title} {Clustering without geometry in sparse networks with independent edges},\ }\bibfield  {journal} {\bibinfo  {journal} {arXiv}\ }\href {https://doi.org/10.48550/arXiv.2603.13159} {10.48550/arXiv.2603.13159} (\bibinfo {year} {2026}{\natexlab{a}})\BibitemShut {NoStop}%
\bibitem [{\citenamefont {Catanzaro}\ \emph {et~al.}(2026{\natexlab{b}})\citenamefont {Catanzaro}, \citenamefont {Hazra},\ and\ \citenamefont {Garlaschelli}}]{catanzaro2025spectra}%
  \BibitemOpen
  \bibfield  {author} {\bibinfo {author} {\bibfnamefont {A.}~\bibnamefont {Catanzaro}}, \bibinfo {author} {\bibfnamefont {R.~S.}\ \bibnamefont {Hazra}},\ and\ \bibinfo {author} {\bibfnamefont {D.}~\bibnamefont {Garlaschelli}},\ }\bibfield  {title} {\bibinfo {title} {Spectra of random graphs with discrete scale invariance},\ }\bibfield  {journal} {\bibinfo  {journal} {npj Complexity}\ }\href {https://doi.org/10.1038/s44260-026-00098-8} {10.1038/s44260-026-00098-8} (\bibinfo {year} {2026}{\natexlab{b}})\BibitemShut {NoStop}%
\bibitem [{\citenamefont {Avena}\ \emph {et~al.}(2026{\natexlab{b}})\citenamefont {Avena}, \citenamefont {Garlaschelli}, \citenamefont {Hazra},\ and\ \citenamefont {den Hollander}}]{avena2026renormalisation}%
  \BibitemOpen
  \bibfield  {author} {\bibinfo {author} {\bibfnamefont {L.}~\bibnamefont {Avena}}, \bibinfo {author} {\bibfnamefont {D.}~\bibnamefont {Garlaschelli}}, \bibinfo {author} {\bibfnamefont {R.~S.}\ \bibnamefont {Hazra}},\ and\ \bibinfo {author} {\bibfnamefont {F.}~\bibnamefont {den Hollander}},\ }\bibfield  {title} {\bibinfo {title} {Renormalisation of inhomogeneous random graphs},\ }\href@noop {} {\bibfield  {journal} {\bibinfo  {journal} {arXiv preprint arXiv:2607.12459}\ } (\bibinfo {year} {2026}{\natexlab{b}})}\BibitemShut {NoStop}%
\bibitem [{\citenamefont {Chen}\ \emph {et~al.}(2023)\citenamefont {Chen}, \citenamefont {Su},\ and\ \citenamefont {Zeng}}]{Chen2023WeightedGR}%
  \BibitemOpen
  \bibfield  {author} {\bibinfo {author} {\bibfnamefont {D.}~\bibnamefont {Chen}}, \bibinfo {author} {\bibfnamefont {H.}~\bibnamefont {Su}},\ and\ \bibinfo {author} {\bibfnamefont {Z.}~\bibnamefont {Zeng}},\ }\bibfield  {title} {\bibinfo {title} {Geometric renormalization reveals the self-similarity of weighted networks},\ }\href {https://doi.org/10.1109/TCSS.2022.3164975} {\bibfield  {journal} {\bibinfo  {journal} {IEEE Transactions on Computational Social Systems}\ }\textbf {\bibinfo {volume} {10}},\ \bibinfo {pages} {426} (\bibinfo {year} {2023})}\BibitemShut {NoStop}%
\bibitem [{\citenamefont {Zheng}\ \emph {et~al.}(2024)\citenamefont {Zheng}, \citenamefont {Garc{\'i}a-P{\'e}rez}, \citenamefont {Bogu{\~n}{\'a}},\ and\ \citenamefont {Serrano}}]{Zheng2024WeightedGR}%
  \BibitemOpen
  \bibfield  {author} {\bibinfo {author} {\bibfnamefont {M.}~\bibnamefont {Zheng}}, \bibinfo {author} {\bibfnamefont {G.}~\bibnamefont {Garc{\'i}a-P{\'e}rez}}, \bibinfo {author} {\bibfnamefont {M.}~\bibnamefont {Bogu{\~n}{\'a}}},\ and\ \bibinfo {author} {\bibfnamefont {M.~{\'A}.}\ \bibnamefont {Serrano}},\ }\bibfield  {title} {\bibinfo {title} {Geometric renormalization of weighted networks},\ }\href {https://doi.org/10.1038/s42005-024-01589-7} {\bibfield  {journal} {\bibinfo  {journal} {Communications Physics}\ }\textbf {\bibinfo {volume} {7}},\ \bibinfo {pages} {97} (\bibinfo {year} {2024})}\BibitemShut {NoStop}%
\bibitem [{\citenamefont {Caldarelli}\ \emph {et~al.}(2024)\citenamefont {Caldarelli}, \citenamefont {Gabrielli}, \citenamefont {Gili},\ and\ \citenamefont {Villegas}}]{Caldarelli2024LaplacianRG}%
  \BibitemOpen
  \bibfield  {author} {\bibinfo {author} {\bibfnamefont {G.}~\bibnamefont {Caldarelli}}, \bibinfo {author} {\bibfnamefont {A.}~\bibnamefont {Gabrielli}}, \bibinfo {author} {\bibfnamefont {T.}~\bibnamefont {Gili}},\ and\ \bibinfo {author} {\bibfnamefont {P.}~\bibnamefont {Villegas}},\ }\bibfield  {title} {\bibinfo {title} {Laplacian renormalization group: an introduction to heterogeneous coarse-graining},\ }\href {https://doi.org/10.1088/1742-5468/ad57b1} {\bibfield  {journal} {\bibinfo  {journal} {Journal of Statistical Mechanics: Theory and Experiment}\ }\textbf {\bibinfo {volume} {2024}},\ \bibinfo {pages} {084002} (\bibinfo {year} {2024})}\BibitemShut {NoStop}%
\bibitem [{\citenamefont {Marzi}\ \emph {et~al.}(2026{\natexlab{a}})\citenamefont {Marzi}, \citenamefont {Pijpers},\ and\ \citenamefont {Garlaschelli}}]{Marzi2026WeightedNetworksAcrossScales}%
  \BibitemOpen
  \bibfield  {author} {\bibinfo {author} {\bibfnamefont {M.}~\bibnamefont {Marzi}}, \bibinfo {author} {\bibfnamefont {F.~P.}\ \bibnamefont {Pijpers}},\ and\ \bibinfo {author} {\bibfnamefont {D.}~\bibnamefont {Garlaschelli}},\ }\bibfield  {title} {\bibinfo {title} {Multiscale reconstruction of weighted networks from coarse-grained data},\ }\bibfield  {journal} {\bibinfo  {journal} {arXiv}\ }\href {https://doi.org/10.48550/arXiv.2609.37957} {10.48550/arXiv.2609.37957} (\bibinfo {year} {2026}{\natexlab{a}})\BibitemShut {NoStop}%
\bibitem [{\citenamefont {Feller}(1968)}]{Feller1968}%
  \BibitemOpen
  \bibfield  {author} {\bibinfo {author} {\bibfnamefont {W.}~\bibnamefont {Feller}},\ }\href@noop {} {\emph {\bibinfo {title} {An Introduction to Probability Theory and Its Applications}}},\ \bibinfo {edition} {3rd}\ ed.,\ Vol.~\bibinfo {volume} {1}\ (\bibinfo  {publisher} {John Wiley \& Sons},\ \bibinfo {year} {1968})\BibitemShut {NoStop}%
\bibitem [{\citenamefont {Ospina}\ and\ \citenamefont {Gerber}(1987)}]{OspinaGerber1987}%
  \BibitemOpen
  \bibfield  {author} {\bibinfo {author} {\bibfnamefont {A.~V.}\ \bibnamefont {Ospina}}\ and\ \bibinfo {author} {\bibfnamefont {H.~U.}\ \bibnamefont {Gerber}},\ }\bibfield  {title} {\bibinfo {title} {A simple proof of {Feller}'s characterization of the compound {Poisson} distributions},\ }\href {https://doi.org/10.1016/0167-6687(87)90008-4} {\bibfield  {journal} {\bibinfo  {journal} {Insurance: Mathematics and Economics}\ }\textbf {\bibinfo {volume} {6}},\ \bibinfo {pages} {63} (\bibinfo {year} {1987})}\BibitemShut {NoStop}%
\bibitem [{\citenamefont {Feller}(1971)}]{Feller1971}%
  \BibitemOpen
  \bibfield  {author} {\bibinfo {author} {\bibfnamefont {W.}~\bibnamefont {Feller}},\ }\href@noop {} {\emph {\bibinfo {title} {An Introduction to Probability Theory and Its Applications, Volume II}}},\ \bibinfo {edition} {2nd}\ ed.\ (\bibinfo  {publisher} {John Wiley \& Sons},\ \bibinfo {address} {New York},\ \bibinfo {year} {1971})\BibitemShut {NoStop}%
\bibitem [{\citenamefont {Johnson}\ \emph {et~al.}(2005)\citenamefont {Johnson}, \citenamefont {Kemp},\ and\ \citenamefont {Kotz}}]{JohnsonKempKotz2005}%
  \BibitemOpen
  \bibfield  {author} {\bibinfo {author} {\bibfnamefont {N.~L.}\ \bibnamefont {Johnson}}, \bibinfo {author} {\bibfnamefont {A.~W.}\ \bibnamefont {Kemp}},\ and\ \bibinfo {author} {\bibfnamefont {S.}~\bibnamefont {Kotz}},\ }\href {https://doi.org/10.1002/0471715816} {\emph {\bibinfo {title} {Univariate Discrete Distributions}}},\ \bibinfo {edition} {3rd}\ ed.\ (\bibinfo  {publisher} {John Wiley \& Sons},\ \bibinfo {address} {Hoboken, NJ},\ \bibinfo {year} {2005})\BibitemShut {NoStop}%
\bibitem [{\citenamefont {Ialongo}\ \emph {et~al.}(2022)\citenamefont {Ialongo}, \citenamefont {de~Valk}, \citenamefont {Marchese}, \citenamefont {Jansen}, \citenamefont {Zmarrou}, \citenamefont {Squartini},\ and\ \citenamefont {Garlaschelli}}]{ReconstructingFirmLevelInteractionsDutchInputOutputNetwork2022}%
  \BibitemOpen
  \bibfield  {author} {\bibinfo {author} {\bibfnamefont {L.~N.}\ \bibnamefont {Ialongo}}, \bibinfo {author} {\bibfnamefont {C.}~\bibnamefont {de~Valk}}, \bibinfo {author} {\bibfnamefont {E.}~\bibnamefont {Marchese}}, \bibinfo {author} {\bibfnamefont {F.}~\bibnamefont {Jansen}}, \bibinfo {author} {\bibfnamefont {H.}~\bibnamefont {Zmarrou}}, \bibinfo {author} {\bibfnamefont {T.}~\bibnamefont {Squartini}},\ and\ \bibinfo {author} {\bibfnamefont {D.}~\bibnamefont {Garlaschelli}},\ }\bibfield  {title} {\bibinfo {title} {Reconstructing firm-level interactions in the dutch input--output network from production constraints},\ }\href {https://doi.org/10.1038/s41598-022-15714-4} {\bibfield  {journal} {\bibinfo  {journal} {Scientific Reports}\ }\textbf {\bibinfo {volume} {12}},\ \bibinfo {pages} {11847} (\bibinfo {year} {2022})}\BibitemShut {NoStop}%
\bibitem [{\citenamefont {Parisi}\ \emph {et~al.}(2020)\citenamefont {Parisi}, \citenamefont {Squartini},\ and\ \citenamefont {Garlaschelli}}]{CReM2020}%
  \BibitemOpen
  \bibfield  {author} {\bibinfo {author} {\bibfnamefont {F.}~\bibnamefont {Parisi}}, \bibinfo {author} {\bibfnamefont {T.}~\bibnamefont {Squartini}},\ and\ \bibinfo {author} {\bibfnamefont {D.}~\bibnamefont {Garlaschelli}},\ }\bibfield  {title} {\bibinfo {title} {A faster horse on a safer trail: Generalised inference for the efficient reconstruction of weighted networks},\ }\href {https://doi.org/10.1088/1367-2630/ab74a7} {\bibfield  {journal} {\bibinfo  {journal} {New Journal of Physics}\ }\textbf {\bibinfo {volume} {22}},\ \bibinfo {pages} {053053} (\bibinfo {year} {2020})}\BibitemShut {NoStop}%
\bibitem [{\citenamefont {Marzi}\ \emph {et~al.}(2026{\natexlab{b}})\citenamefont {Marzi}, \citenamefont {Giuffrida}, \citenamefont {Garlaschelli},\ and\ \citenamefont {Squartini}}]{marzi2026reproducing}%
  \BibitemOpen
  \bibfield  {author} {\bibinfo {author} {\bibfnamefont {M.}~\bibnamefont {Marzi}}, \bibinfo {author} {\bibfnamefont {F.}~\bibnamefont {Giuffrida}}, \bibinfo {author} {\bibfnamefont {D.}~\bibnamefont {Garlaschelli}},\ and\ \bibinfo {author} {\bibfnamefont {T.}~\bibnamefont {Squartini}},\ }\bibfield  {title} {\bibinfo {title} {Reproducing the first and second moments of empirical degree distributions},\ }\href {https://doi.org/10.1103/3vtj-5nlt} {\bibfield  {journal} {\bibinfo  {journal} {Phys.\ Rev.\ Res.}\ }\textbf {\bibinfo {volume} {8}},\ \bibinfo {pages} {013141} (\bibinfo {year} {2026}{\natexlab{b}})}\BibitemShut {NoStop}%
\bibitem [{\citenamefont {Marzi}\ and\ \citenamefont {Squartini}(2026)}]{marzi2026Bayesian}%
  \BibitemOpen
  \bibfield  {author} {\bibinfo {author} {\bibfnamefont {M.}~\bibnamefont {Marzi}}\ and\ \bibinfo {author} {\bibfnamefont {T.}~\bibnamefont {Squartini}},\ }\bibfield  {title} {\bibinfo {title} {A bayesian approach to out-of-sample network reconstruction},\ }\href {https://doi.org/10.1038/s42005-026-02725-1} {\bibfield  {journal} {\bibinfo  {journal} {Commun. Phys.}\ }\textbf {\bibinfo {volume} {9}},\ \bibinfo {pages} {316} (\bibinfo {year} {2026})}\BibitemShut {NoStop}%
\bibitem [{\citenamefont {Verteletskyi}(2022)}]{Verteletskyi2022}%
  \BibitemOpen
  \bibfield  {author} {\bibinfo {author} {\bibfnamefont {V.}~\bibnamefont {Verteletskyi}},\ }\emph {\bibinfo {title} {Renormalization of networks with weighted links}},\ \href {https://hdl.handle.net/1887/3453824} {Master's thesis},\ \bibinfo  {school} {Leiden University} (\bibinfo {year} {2022})\BibitemShut {NoStop}%
\bibitem [{\citenamefont {Sundt}(2000)}]{Sundt2000}%
  \BibitemOpen
  \bibfield  {author} {\bibinfo {author} {\bibfnamefont {B.}~\bibnamefont {Sundt}},\ }\bibfield  {title} {\bibinfo {title} {Multivariate compound {Poisson} distributions and infinite divisibility},\ }\href {https://doi.org/10.2143/AST.30.2.504637} {\bibfield  {journal} {\bibinfo  {journal} {ASTIN Bulletin}\ }\textbf {\bibinfo {volume} {30}},\ \bibinfo {pages} {305} (\bibinfo {year} {2000})}\BibitemShut {NoStop}%
\end{thebibliography}%





\clearpage



%
%
\onecolumngrid
\setcounter{equation}{0}

\setcounter{figure}{0}
\setcounter{table}{0}
\setcounter{page}{1}
\setcounter{section}{0}

\renewcommand{\theequation}{S\arabic{equation}}
\renewcommand{\thetable}{S\arabic{table}}
\renewcommand{\thefigure}{S\arabic{figure}}
\renewcommand{\thesection}{S.\Roman{section}} 
\renewcommand{\thesubsection}{\thesection.\roman{subsection}}

\makeatletter
\renewcommand{\p@subsection}{}
\makeatother

\vspace*{2.5em}
\begin{center}
    \definecolor{supplementink}{RGB}{34,55,79}
    {\fontsize{14.4}{19}\selectfont\mdseries\scshape\color{supplementink} Supplementary Information\par}
    \vspace{0.9em}
    {\small accompanying the paper\par}
    \vspace{0.35em}
    {\fontsize{25}{31}\selectfont\bfseries
        Multiscale Renormalization\\[0.16em]
        of Weighted Networks\par}
    \vspace{0.75em}
    {\color{supplementink}\rule{0.68\textwidth}{0.7pt}}\par
    \vspace{0.7em}
    {\large Mattia Marzi\textsuperscript{1,2,3,4,*},\enspace
    Frank P. Pijpers\textsuperscript{3,5},\enspace
    and Diego Garlaschelli\textsuperscript{1,2,4}\par}
    \vspace{0.65em}
    \begin{minipage}{0.94\textwidth}
        \centering\footnotesize
        \textsuperscript{1}IMT School for Advanced Studies, P.zza San Francesco 19, 55100 Lucca (Italy)\par
        \textsuperscript{2}Lorentz Institute for Theoretical Physics, University of Leiden, Einsteinweg 55, 2333 CC Leiden (The Netherlands)\par
        \textsuperscript{3}Statistics Netherlands, Henri Faasdreef 312, 2492 JP Den Haag (the Netherlands)\par
        \textsuperscript{4}INdAM-GNAMPA Istituto Nazionale di Alta Matematica `Francesco Severi', P.le Aldo Moro 5, 00185 Rome (Italy)\par
        \textsuperscript{5}Korteweg - de Vries Institute for Mathematics, University of Amsterdam, Amsterdam (the Netherlands)\par
        \vspace{0.55em}
        \textsuperscript{*}Corresponding author
        \href{mailto:mattia.marzi@imtlucca.it}{\nolinkurl{mattia.marzi@imtlucca.it}}\par
    \end{minipage}\par
    \vspace{2.0em}
    \begin{minipage}{0.81\textwidth}
        \hypersetup{linkcolor=supplementink,linktoc=all}
        \setcounter{tocdepth}{2}
        \makeatletter
        \addtocontents{toc}{\protect\tocdepth@restore}
        \makeatother
        \tableofcontents
    \end{minipage}
\end{center}
\clearpage

\section{The binary MultiScale Model}
\hypertarget{AppA}{}

Many real networks admit natural hierarchies of aggregation. Multiscale network renormalization
identifies random graph ensembles whose functional form is preserved under arbitrary coarse
graining~\cite{Garuccio2023,LalliGarlaschelli2024,Gabrielli2025}. Here we summarize the derivation of the independent-dyad
binary family compatible with additive parameter renormalization in the undirected case.

Consider a microscopic undirected network with adjacency matrix
$A^{(0)}=\{a^{(0)}_{ij}\}$ and independent dyads. Let nodes be partitioned into disjoint blocks
$I,J$. For two distinct blocks, $I\neq J$, logical OR coarse graining gives
\begin{equation}
a^{(1)}_{IJ}=1-\prod_{i\in I}\prod_{j\in J}\left(1-a^{(0)}_{ij}\right),\qquad{I\neq J},
\end{equation}
so that the induced probability satisfies
\begin{equation}
p_{IJ}=1-\prod_{i\in I}\prod_{j\in J}\left(1-p_{ij}\right),\qquad{I\neq J}.
\end{equation}
Scale invariance requires a microscopic family
\begin{equation}
p_{ij}=f(x_i,x_j),\qquad{i\neq j},
\end{equation}
with additive renormalization
\begin{equation}
x_I=\sum_{i\in I}x_i,
\end{equation}
such that
\begin{equation}
p_{IJ}=f(x_I,x_J),\qquad{I\neq J}
\end{equation}
for every pair of distinct blocks. Define $g(u,v)=\log(1-f(u,v))$. The invariance condition becomes
\begin{equation}
g(x_I,x_J)=\sum_{i\in I}\sum_{j\in J}g(x_i,x_j),\qquad{I\neq J}.
\end{equation}
Assuming the minimal product ansatz $g(u,v)=h(uv)$ yields
\begin{equation}
h(x_I x_J)=\sum_{i\in I}\sum_{j\in J}h(x_i x_j),\qquad{I\neq J}.
\end{equation}
For $I=\{i_1,i_2\}$ and $J=\{j\}$ this reduces to $h(z_1+z_2)=h(z_1)+h(z_2)$ with
$z_k=x_{i_k}x_j$. Regular solutions are linear, $h(z)=cz$. Re-parametrizing the negative constant $c$ gives
\begin{equation}
\label{eq:sim}
p_{ij}=1-\exp(-\delta x_i x_j),\qquad{i\neq j}
\end{equation}
with $\delta>0$, which defines the undirected binary MSM~\cite{Garuccio2023,LalliGarlaschelli2024,Gabrielli2025}.
The exponential bilinear structure is therefore the independent-dyad specification whose
functional form commutes with aggregation between distinct blocks.

{
\subsection{Diagonal binary probabilities}
\label{sec:binary_diagonal}
A block has no self-loop if it contains neither a microscopic self-loop nor an edge between distinct internal nodes. Under independence of unordered dyads, the OR rule is~\cite{avena2026renormalisation}
\begin{equation}
1-p_{II}=\prod_{i\in I}(1-p_{ii})\prod_{i\in I}\prod_{\substack{j\in I\\j>i}}(1-p_{ij}).
\label{eq:binary_loop_product}
\end{equation}
Each internal edge is counted once in this binary product. Write $d(x)=-\log(1-p_{ii})$ for a node with field $x$. Merging two nodes with fields $u,v$ and using the off-diagonal law gives
\begin{equation}
d(u+v)=d(u)+d(v)+\delta uv.
\end{equation}
Subtracting the quadratic term makes the remainder additive:
\begin{equation}
r(x)=d(x)-\frac{\delta x^2}{2},\qquad r(u+v)=r(u)+r(v).
\end{equation}
Under continuity, $r(x)=\eta x/2$, with a global invariant parameter $\eta$. Hence~\cite{MiloccoJansenGarlaschelli2024,avena2026renormalisation}
\begin{equation}
\label{eq:diag}
p_{ii}=1-\exp\!\left[-\frac{\delta x_i^2+\eta x_i}{2}\right],\qquad i=j.
\end{equation}
Substitution into Eq.~\eqref{eq:binary_loop_product} yields
\begin{equation}
-\log(1-p_{II})=\frac{\delta}{2}\sum_{i\in I}x_i^2+\frac{\eta}{2}\sum_{i\in I}x_i
+\delta\sum_{i\in I}\sum_{\substack{j\in I\\j>i}}x_ix_j=\frac{\delta x_I^2+\eta x_I}{2}.
\end{equation}
For all positive fields to be allowed, $\eta\ge0$ is required. On a fixed finite microscopic field set, binary admissibility instead requires $\eta\ge-\delta x_{\min}$, where $x_{\min}=\min_{i:x_i>0}x_i$; nodes with zero field impose no additional condition. This microscopic bound also protects aggregate fields. Binary admissibility alone does not guarantee a valid weighted diagonal law, as discussed in Sec.~\ref{sec:weighted_diagonal}.
}

\subsection{{Strength-based parametrization}}

In many applications (e.g. economic networks), degrees are not observed while node strengths are available.
A natural fitness-based specification therefore ties the node fields directly to strengths. Setting
\begin{equation}
x_i=s_i,
\end{equation}
one obtains
\begin{equation}
p_{ij}=1-\exp(-\delta s_i s_j),\qquad{i\neq j},
\end{equation}
where $\delta>0$ is a single global parameter controlling overall density. This is what we call the binary MSM. Since strengths renormalize
additively, $s_I=\sum_{i\in I}s_i$, the same functional form holds at every aggregation level for inter-block dyads.

\subsection{{Dyadic frictions}}

Starting from eq.~\ref{eq:sim}, one can introduce dyadic frictions without breaking scale invariance by
considering
\begin{equation}
p_{ij}=1-\exp(-\delta x_i x_j f_{ij}),\qquad{i\neq j},
\label{eq:sim_dyadic}
\end{equation}
where $f_{ij}\ge0$ is a pair-specific factor. Under logical OR coarse graining and dyadic independence,
\begin{equation}
1-p_{IJ}=\prod_{i\in I}\prod_{j\in J}(1-p_{ij})
=\exp\!\left(-\delta\sum_{i\in I}\sum_{j\in J}x_i x_j f_{ij}\right),\qquad{I\neq J}.
\label{eq:or_product_dyadic}
\end{equation}
If node fields renormalize additively, $x_I=\sum_{i\in I}x_i$, then the block
probability can be written in the same functional form as the microscopic one provided the dyadic factor
renormalizes as
\begin{equation}
f_{IJ}=\frac{\sum_{i\in I}\sum_{j\in J}x_i x_j f_{ij}}{\left(\sum_{i\in I}x_i\right)\left(\sum_{j\in J}x_j\right)},\qquad{I\neq J}.
\label{eq:dyadic_renorm}
\end{equation}
With this definition one obtains $p_{IJ}=1-\exp(-\delta x_I x_J f_{IJ})$, hence exact closure under arbitrary coarse graining between distinct blocks.

\clearpage


\section{The weighted MultiScale Model}
\hypertarget{AppB}{}

Here we provide our detailed derivation of the aggregation-invariant network ensemble for integer-valued weights $w_{ij}\in\mathbb N$ that is closed under coarse graining by summation {between distinct blocks},
\begin{equation}
w_{IJ}=\sum_{i\in I}\sum_{j\in J}w_{ij},
\qquad I\neq J.
\end{equation}
For diagonal blocks, {$w_{II}=\sum_{i\in I}w_{ii}+2\sum_{i\in I}\sum_{j\in I,\,j>i}w_{ij}$. The diagonal construction is derived separately in Sec.~\ref{sec:weighted_diagonal}.} Related weighted-renormalization ideas for continuous links under more specific assumptions are discussed in~\cite{Verteletskyi2022}. Here we work with discrete weights in a general multiscale setting.

\subsection{PGF factorization and the scale-invariant form}

For nonnegative integer weights $w_{ij}\in\mathbb N$, define the PGF
\begin{equation}
G_{ij}(z)=E[z^{w_{ij}}]=\sum_{n=0}^{\infty}P(w_{ij}=n)\,z^n,\qquad z\in[0,1].
\end{equation}
The PGF uniquely determines the probability mass function (PMF). In particular,
\begin{equation}
P(w_{ij}=n)=\frac{1}{n!}\frac{d^n}{dz^n}G_{ij}(z)\Big|_{z=0},\qquad n\in\mathbb N.
\end{equation}
Normalization follows by evaluating the PGF at $z=1$,
\begin{equation}
G_{ij}(1)=E[1^{w_{ij}}]=\sum_{n=0}^{\infty}P(w_{ij}=n)=1.
\end{equation}
Moreover,
\begin{equation}
P(w_{ij}=0)=G_{ij}(0),
\end{equation}
so the induced binary link probability is
\begin{equation}
p_{ij}=P(w_{ij}>0)=1-P(w_{ij}=0)=1-G_{ij}(0).
\end{equation}
Hence, the binary projection of any weighted ensemble is entirely determined by the value of its PGF at zero.

Assume conditional dyadic independence given node-level fields. Then aggregation by summation implies
\begin{equation}
G_{IJ}(z)=E[z^{w_{IJ}}]
=\prod_{i\in I}\prod_{j\in J}G_{ij}(z),
\qquad I\neq J.
\end{equation}
Scale invariance requires the existence of additive node fields $x_i$ such that
\begin{equation}
x_I=\sum_{i\in I}x_i,
\end{equation}
and the PGF preserves its functional form under aggregation between distinct blocks.

Taking logarithms yields an additivity equation analogous to the binary case,
\begin{equation}
\log G_{IJ}(z)=\sum_{i\in I}\sum_{j\in J}\log G_{ij}(z),\qquad{I\neq J}.
\end{equation}
Repeating the same two-node merge argument used in SI and assuming mild regularity implies that for each fixed $z$ there exists a nonnegative function $\psi(z)$ such that
\begin{equation}
G_{ij}(z)=\exp\!\big(-x_i x_j\,\psi(z)\big),\qquad{i\neq j}.
\end{equation}
Normalization forces $\psi(1)=0$, since $G_{ij}(1)=\exp(-x_i x_j\psi(1))=1$ for all dyads with $x_i x_j>0$.

The mass at zero is therefore
\begin{equation}
P(w_{ij}=0)=\exp\!\big(-x_i x_j\,\psi(0)\big),\qquad{i\neq j}.
\end{equation}
Defining $\delta:=\psi(0)\ge0$ gives the binary projection
\begin{equation}
p_{ij}=1-\exp(-\delta\,x_i x_j),\qquad{i\neq j},
\end{equation}
which coincides exactly with the binary MSM {for $i\neq j$. The diagonal binary projection must be derived from its own PGF}.
{
For $F_a(z)=\exp[-a\psi(z)]$ to be a valid PGF for every $a\ge0$, it must satisfy
\begin{equation}
F_a(z)=\left[F_{a/m}(z)\right]^m
\end{equation}
for every positive integer $m$. The corresponding weight is therefore a sum of $m$ independent, identically distributed nonnegative integer-valued variables for arbitrary $m$. Feller's characterization identifies this infinite divisibility with the compound-Poisson class~\cite{Feller1968,OspinaGerber1987}; see~\cite{Sundt2000} for its multivariate extension. Thus, under this requirement, the compound-Poisson representation is general, whereas the geometric mark law is a separate modeling choice.
}

\subsection{The Poisson case and intrinsic coupling}

A particularly simple admissible specification of the function $\psi(z)$ in the exponential PGF is the linear form $\psi(z)=\delta(1-z)$, with $\delta\ge0$, which yields
\begin{equation}
G_{ij}(z)=\exp\!\big(-\delta\,x_i x_j(1-z)\big),\qquad{i\neq j}.
\end{equation}

Notice that the function $\exp(\lambda(z-1))$ is the probability generating function of a Poisson random variable with parameter $\lambda$, since
\begin{equation}
\exp(\lambda(z-1))=e^{-\lambda}\sum_{n=0}^{\infty}\frac{\lambda^n}{n!}z^n,
\end{equation}
which implies
\begin{equation}
P(W=n)=e^{-\lambda}\frac{\lambda^n}{n!},\qquad n\in\mathbb N.
\end{equation}
In our case, we define
\begin{equation}
\lambda_{ij}:=\delta\,x_i x_j,\qquad{i\neq j},
\end{equation}
and the PGF therefore reads
\begin{equation}
G_{ij}(z)=\exp\!\big(\lambda_{ij}(z-1)\big),
\end{equation}
so that
\begin{equation}
w_{ij}\sim\mathrm{Poisson}(\lambda_{ij}).
\end{equation}
The induced binary projection and the expected weight are
\begin{equation}
p_{ij}=1-\exp(-\lambda_{ij}),
\qquad
E[w_{ij}]=\lambda_{ij}.
\end{equation}
Hence, the same parameter $\lambda_{ij}$ controls both topology and weight magnitude. Once the binary layer is fixed, the weighted layer is mechanically pinned down through
\begin{equation}
E[w_{ij}]=-\log(1-p_{ij}).
\end{equation}
This rigidity is undesirable in reconstruction settings where link density and aggregate weight constraints originate from distinct sources of information.

\subsection{Compound-Poisson construction and separation of layers}

To preserve scale invariance while introducing the minimal additional degree of freedom required for reconstruction, we adopt a compound-Poisson specification. For each pair $(i,j)$ define
\begin{equation}
w_{ij}=\sum_{r=1}^{K_{ij}}X_{ij,r},
\qquad
K_{ij}\sim\mathrm{Poisson}(\lambda_{ij}),
\end{equation}
where $\lambda_{ij}\ge0$ is the dyadic intensity and the marks $X_{ij,r}\in\{1,2,\dots\}$ are identically distributed and mutually independent, and are also independent of $K_{ij}$. Let $X$ denote a generic mark with this common distribution and define its probability generating function
\begin{equation}
h(z)=E[z^X].
\end{equation}

By the law of total expectation,
\begin{equation}
G_{ij}(z)=E[z^{w_{ij}}]=E\!\left[E[z^{w_{ij}}\mid K_{ij}]\right].
\end{equation}
Given $K_{ij}=k$, the dyadic weight is the sum of $k$ i.i.d.\ marks, hence
\begin{equation}
E[z^{w_{ij}}\mid K_{ij}=k]
=E\!\left[z^{\sum_{r=1}^{k}X_{ij,r}}\right]
=\prod_{r=1}^{k}E[z^{X_{ij,r}}]
=(h(z))^{k}.
\end{equation}
Substituting into the outer expectation gives
\begin{equation}
G_{ij}(z)=E[(h(z))^{K_{ij}}]
=\sum_{k=0}^{\infty}(h(z))^{k}P(K_{ij}=k).
\end{equation}
Since $K_{ij}\sim\mathrm{Poisson}(\lambda_{ij})$, one has
\begin{equation}
P(K_{ij}=k)=e^{-\lambda_{ij}}\frac{\lambda_{ij}^k}{k!},
\end{equation}
hence
\begin{equation}
G_{ij}(z)
=e^{-\lambda_{ij}}\sum_{k=0}^{\infty}
\frac{(\lambda_{ij}h(z))^{k}}{k!}
=e^{-\lambda_{ij}}\,\exp\!\big(\lambda_{ij}h(z)\big)
=\exp\!\Big(\lambda_{ij}(h(z)-1)\Big).
\end{equation}

Since $X\ge1$ implies $h(0)=0$, the mass at zero becomes
\begin{equation}
P(w_{ij}=0)=G_{ij}(0)=\exp(-\lambda_{ij}),
\qquad
p_{ij}=1-\exp(-\lambda_{ij}),
\end{equation}
which depends only on $\lambda_{ij}$ and is independent of the mark distribution.

The expected weight follows from iterated expectation,
\begin{align}
E[w_{ij}]
&=E\!\left[E\!\left[\sum_{r=1}^{K_{ij}}X_{ij,r}\,\Big|\,K_{ij}\right]\right]
=E\!\left[K_{ij}E[X]\right]
=\lambda_{ij}E[X].
\end{align}
Thus $\lambda_{ij}$ controls the binary projection, while the mark mean $E[X]$ rescales positive weights without affecting $p_{ij}$. This explicit separation is precisely the additional degree of freedom absent in the pure Poisson case.

At this stage the construction is still incomplete, since the mark distribution has not yet been specified. To obtain a fully explicit and analytically tractable model, we choose the geometric distribution on $\{1,2,\dots\}$,
\begin{equation}
P(X=n)=\rho(1-\rho)^{n-1},\qquad n\ge1,
\end{equation}
with mean $E[X]=1/\rho${, for $0<\rho\le1$. This mark law maximizes Shannon entropy on the positive integers at a fixed finite mean}. Then
\begin{equation}
E[w_{ij}]=\frac{\lambda_{ij}}{\rho}.
\end{equation}
The PMF follows by conditioning on the number of marks,
\begin{equation}
P(w_{ij}=n)=\sum_{k=0}^{\infty}P(K_{ij}=k)\,P\!\left(\sum_{r=1}^{k}X_{ij,r}=n\right)=\sum_{k=0}^{\infty}e^{-\lambda_{ij}}\frac{\lambda_{ij}^k}{k!}\,P\!\left(\sum_{r=1}^{k}X_{ij,r}=n\right).
\end{equation}
For geometric marks on $\{1,2,\dots\}$, the sum of $k$ i.i.d.\ variables has negative-binomial distribution,
\begin{equation}
P\!\left(\sum_{r=1}^{k}X_{ij,r}=n\right)
=\binom{n-1}{k-1}\rho^k(1-\rho)^{\,n-k},\qquad n\ge k\ge1.
\end{equation}
Since $w_{ij}=0$ only when $K_{ij}=0$, one obtains
\begin{equation}
P(w_{ij}=0)=e^{-\lambda_{ij}},
\end{equation}
and for $n\ge1$
\begin{equation}
P(w_{ij}=n)=e^{-\lambda_{ij}}
\sum_{k=1}^{n}
\frac{\lambda_{ij}^k}{k!}
\binom{n-1}{k-1}
\rho^k(1-\rho)^{\,n-k}.
\end{equation}

Closure under coarse graining {between distinct blocks} follows from additivity of Poisson intensities. A sum of independent compound-Poisson variables with the same mark law is again compound{-}Poisson with intensity equal to the sum of intensities and unchanged mark distribution.

\subsection{Aggregation-invariant weighted self-loops}
\label{sec:weighted_diagonal}
For an undirected weighted matrix, the diagonal of a block is
\begin{equation}
w_{II}=\sum_{i\in I}w_{ii}+2\sum_{\substack{i,j\in I\\i<j}}w_{ij}.
\label{eq:weighted_block_diagonal}
\end{equation}
The factor two counts both symmetric entries of each internal edge and ensures $s_I=\sum_{i\in I}s_i$. Conditional independence therefore gives
\begin{equation}
G_{II}(z)=\prod_{i\in I}G_{ii}(z)\prod_{\substack{i,j\in I\\i<j}}G_{ij}(z^2).
\label{eq:weighted_loop_product}
\end{equation}
The argument $z^2$ accounts for the doubled weights.

With $G_{ij}(z)=\exp[-x_ix_j\psi(z)]$ for $i\ne j$, write $\ell(x,z)=-\log G_{ii}(z)$ for a node with field $x$. Merging fields $u,v$ requires
\begin{equation}
\ell(u+v,z)=\ell(u,z)+\ell(v,z)+uv\psi(z^2).
\end{equation}
Since $(u+v)^2-u^2-v^2=2uv$, the merging condition implies that $\ell(x,z)-x^2\psi(z^2)/2$ is additive in $x$. Under continuity, this difference is proportional to $x$, so $\ell(x,z)=x^2\psi(z^2)/2+x\chi(z)/2$, where $\chi(z)$ is independent of $x$. Hence,
\begin{equation}
G_{ii}(z)=\exp\!\left[-\frac{x_i^2\psi(z^2)+x_i\chi(z)}{2}\right].
\label{eq:general_weighted_diagonal}
\end{equation}
Normalization requires $\chi(1)=0$, and the binary law~\eqref{eq:diag} requires $\chi(0)=\eta$; validity as a PGF imposes further conditions.

To give the diagonal component its own weight scale, retain $\psi(z)=\delta[1-h(z)]$ and choose $\chi(z)=\eta[1-t(z)]$, with geometric PGFs
\begin{equation}
h(z)=\frac{\rho z}{1-(1-\rho)z},\qquad
t(z)=\frac{\sigma z}{1-(1-\sigma)z}.
\end{equation}
The resulting diagonal law is
\begin{equation}
G_{ii}(z)=\exp\!\bigg\{\frac{\delta x_i^2}{2}[h(z^2)-1]+\frac{\eta x_i}{2}[t(z)-1]\bigg\}.
\label{eq:weighted_diagonal_sigma}
\end{equation}
It combines independent doubled geometric marks of parameter $\rho$ and Poisson intensity $\delta x_i^2/2$ with ordinary geometric marks of parameter $\sigma$ and intensity $\eta x_i/2$. Thus $0<\rho,\sigma\le1$ and $\eta\ge0$ give a valid PGF. Negative $\eta$ is inadmissible, since $P(w_{ii}=1)=G_{ii}(0)\eta x_i\sigma/2<0$ for $x_i>0$. Substitution into Eq.~\eqref{eq:weighted_loop_product} recovers the same law with $x_I=\sum_{i\in I}x_i$: both mark parameters remain invariant.

The binary projection remains Eq.~\eqref{eq:diag}, while
\begin{align}
E[w_{ij}]&=\frac{\delta}{\rho}x_ix_j,\qquad i\ne j,\\
E[w_{ii}]&=\frac{\delta}{\rho}x_i^2+\frac{\eta}{2\sigma}x_i.
\end{align}
Therefore $\sigma$ changes the additional diagonal weight without changing loop probabilities.

Separate calibration first fixes $\delta$ by matching the off-diagonal link count,
\begin{equation}
\sum_{i<j}(1-e^{-\delta x_ix_j})=L_{\mathrm{off}}.
\end{equation}
At fixed $\delta$, the expected loop count
\begin{equation}
C(\eta)=\sum_i\left[1-e^{-(\delta x_i^2+\eta x_i)/2}\right]
\end{equation}
increases strictly with $\eta$. Thus fixing $\eta$ through $C(\eta)=L_{\mathrm{loop}}$ admits a finite admissible solution only if
\begin{equation}
C(0)\le L_{\mathrm{loop}}<M_+,
\label{eq:loop_count_feasibility}
\end{equation}
where $M_+$ is the number of positive-field nodes.

Likewise, $\rho$ is fixed by matching the observed off-diagonal weight $W_{\mathrm{off}}=\sum_{i\ne j}w_{ij}>0$:
\begin{equation}
\sum_{i\ne j}E[w_{ij}]=\frac{\delta}{\rho}\left[\left(\sum_i x_i\right)^2-\sum_i x_i^2\right]=W_{\mathrm{off}},
\end{equation}
giving
\begin{equation}
\rho=\frac{\delta}{W_{\mathrm{off}}}\left[\left(\sum_i x_i\right)^2-\sum_i x_i^2\right].
\end{equation}
At fixed $\delta$, $\eta$ and $\rho$, the expected self-loop weight
\begin{equation}
B(\sigma)=\sum_i E[w_{ii}]=\frac{\delta}{\rho}\sum_i x_i^2+\frac{\eta}{2\sigma}\sum_i x_i
\end{equation}
decreases strictly with $\sigma$ for $\eta>0$. Its minimum over $0<\sigma\le1$ is therefore $B(1)$. Fixing $\sigma$ by matching the observed self-loop weight $W_{\mathrm{loop}}=\sum_i w_{ii}$ through $B(\sigma)=W_{\mathrm{loop}}$ gives
\begin{equation}
\sigma=\frac{\eta\sum_i x_i}{2\left(W_{\mathrm{loop}}-\frac{\delta}{\rho}\sum_i x_i^2\right)}.
\end{equation}
Besides Eq.~\eqref{eq:loop_count_feasibility}, admissibility for $\eta>0$ thus requires
\begin{equation}
0<\rho\le1,\qquad W_{\mathrm{loop}}\ge B(1)=\frac{\delta}{\rho}\sum_i x_i^2+\frac{\eta}{2}\sum_i x_i.
\label{eq:loop_weight_feasibility}
\end{equation}
With $x_i=s_i$, the two weight constraints ensure that the total strength of every node is reproduced in expectation, $E[s_i]=x_i[\delta\sum_j x_j/\rho+\eta/(2\sigma)]=s_i$. The off-diagonal and self-loop weights are matched separately only at the network level, through $W_{\mathrm{off}}$ and $W_{\mathrm{loop}}$; their individual contributions to each node's strength are generally not reproduced. Moreover, these two contributions are not separately additive under aggregation, because internal edges become diagonal block weight.

\subsection{Self-loops in strength-based reconstruction}
\label{sec:empirical_diagonal}
The preceding constraints are frequently infeasible in empirical reconstruction: the off-diagonal fit can already force more self-loops, or more diagonal weight, than observed, and varying $\sigma$ cannot lower either minimum. The companion paper~\cite{Marzi2026WeightedNetworksAcrossScales} therefore adopts a simpler prescription: it assigns the off-diagonal compound-Poisson law also to every retained diagonal, fits the total link count without a separate loop-count constraint, and fixes the weight scale through the observed total strengths.

Specifically, with $x_i=s_i$ and $W=\sum_i s_i$, choosing $\rho=\delta W$ gives
\begin{equation}
E[w_{ij}]=\frac{x_ix_j}{W},\qquad E[s_i]=x_i,
\end{equation}
on full support, provided $0<\rho\le1$. This preserves every total strength and its additivity with a single weight parameter, without imposing separate diagonal counts or weights.

The trade-off is that inter-block distributions remain invariant, whereas diagonal distributions do not. To see this, applying Eq.~\eqref{eq:weighted_loop_product} to the empirical prescription gives
\begin{align}
\log G_{II}^{\mathrm{agg}}(z)&=\delta\sum_{i\in I}x_i^2[h(z)-1]+\frac{\delta}{2}\left(x_I^2-\sum_{i\in I}x_i^2\right)[h(z^2)-1],\\
\log G_{II}^{\mathrm{direct}}(z)&=\delta x_I^2[h(z)-1].
\end{align}
Their derivatives at $z=1$ agree, yielding $E[w_{II}]=\delta x_I^2/\rho$ as required by the gravity expectation, but their zero-weight probabilities and full distributions generally differ. Thus the empirical prescription retains exact inter-block laws and full-matrix means, while relaxing exact diagonal distributional closure.

\end{document}